\documentclass[12pt]{article}

\usepackage[margin=1in]{geometry} 
\usepackage{times} 
\usepackage{setspace} 
\usepackage{graphicx} 
\usepackage{caption} 
\usepackage{subcaption} 
\usepackage{fancyhdr} 
\usepackage{enumitem} 
\usepackage{titlesec} 
\usepackage[hyphens]{url}
\usepackage{hyperref}
\hypersetup{breaklinks=true} 
\usepackage{longtable}
\usepackage[acronym]{glossaries}

\usepackage{float}
\usepackage{multirow}
\usepackage{multicol}
\newcommand{\probP}{\text{I\kern-0.15em P}}
\newcommand{\expec}{\text{I\kern-0.15em E}}
\newcommand{\real}{\text{I\kern-0.15em R}}
\usepackage[ruled, vlined]{algorithm2e}
\usepackage[authoryear]{natbib}
\newacronym{uq}{UQ}{uncertainty quantification}

\newacronym{usaf}{USAF}{United States Air Force}
\newacronym{ussf}{USSF}{United States Space Force}
\newacronym{nhl}{NHL}{Non-Hodgkin Lymphoma}
\newacronym{mafb}{MAFB}{Malmstrom Air Force Base}
\newacronym{mtafb}{$M^T$AFB}{Minot Air Force Base}
\newacronym{or}{OR}{Operations Research}
\newacronym{cdc}{CDC}{Center for Disease Control and Prevention}
\newacronym{ms}{MS}{Missile Squadron}
\newacronym{maf}{MAF}{Missile Alert Facility}
\newacronym{mw}{MW}{Missile Wing}
\newacronym{oss}{OSS}{Operations Support Squadron}
\newacronym{ogv}{OGV}{Operations Group Standardization and Evaluation}
\newacronym{lcc}{LCC}{Launch Control Center}
\newacronym{dlbcl}{DLBCL}{Diffuse Large B-Cell Lymphoma}
\newacronym{mcl}{MCL}{Mantle Cell Lymphoma}
\newacronym{af}{AF}{Air Force}
\newacronym{seer}{SEER}{Surveillance, Epidemiology, and End Results Program}
\newacronym{SD}{sd}{standard deviation}
\newacronym{vsfb}{VSFB}{Vandenberg Space Force Base}
\newacronym{trs}{TRS}{Training Squadron}
\newacronym{ist}{IST}{Initial Skills Training}
\newacronym{us}{US}{United States}
\newacronym{sir}{SIR}{Standardized Incidence Ratio}
\newacronym{pmf}{PMF}{Probability Mass Function}
\newacronym{icbm}{ICBM}{Intercontinental Ballistic Missile}
\newacronym{nci}{NCI}{National Cancer Institute}
\newacronym{acs}{ACS}{American Cancer Society}
\newacronym{hiv}{HIV}{Human Immunodeficiency Virus}
\newacronym{dod}{DoD}{Department of Defense}
\newacronym{pcbs}{PCBs}{Polychlorinated Biphenyls}
\newacronym{cdf}{CDF}{cumulative distribution function}
\makeglossaries

\title{\textbf{ICBM community cancer registry analysis: a focus on Non-Hodgkin Lymphoma cases in missileers}}
\author{
    Dawn L. Sanderson \\
    Bldg 641 Rm 227\\
    2950 Hobson Way\\
    Wright-Patterson AFB, OH 45433\\
    dawn.sanderson@us.af.mil\\
    \\
    Richard L. Smith, Department of Statistics and Operations Research, UNC Chapel Hill
}
\date{}

\begin{document}

\maketitle

\subsection*{Disclaimer}
The views expressed are those of the authors and do not reflect the official guidance or position of the United
States Government, the Department of Defense the United States Air Force or the United States Space Force.
\subsection*{Abstract}
\noindent This study investigates the incidence and age at diagnosis of Non-Hodgkin Lymphoma (NHL) among missileers stationed at Malmstrom Air Force Base (MAFB) compared to national benchmarks. The analysis was motivated by reports of elevated cancer diagnoses within the Intercontinental Ballistic Missile (ICBM) community, specifically targeting NHL cases due to initial media focus and data collection through the Torchlight Initiative. The methodology integrates simulation-based estimation of expected diagnoses using incomplete data and expert knowledge on the underlying population. Statistical tests, including the Standardized Incidence Ratio (SIR) and a non-parametric Sign Test, were used to evaluate both the rate of diagnosis and age at diagnosis. The results demonstrate a statistically significant increase in NHL diagnoses among missileers in the later decades, with observed rates surpassing expected benchmarks. The study also finds that the median age of diagnosis is significantly younger for the study population compared to national averages. Key methodological contributions include estimating the population size and service start ages when comprehensive cohort data is unavailable, incorporating uncertainty quantification, and applying multiple hypothesis testing to identify temporal patterns. While the study acknowledges limitations such as small sample size and estimation uncertainty, as well as recognizing the study subjects as self-reported diagnoses, the findings highlight the effectiveness of these statistical techniques in identifying significant deviations from expected rates. Future studies should refine and build upon these methods, incorporating survival analysis and additional covariates to improve the robustness and scope of the results.

\section{Introduction}\label{intro}

In early 2023, an organization known as the Torchlight Initiative, was established to collect reports of cancer diagnoses and related fatalities from former personnel and family members associated with the \gls{icbm} community \citep{torchlight}. As this data collection initiative progressed, emerging patterns warranted a detailed statistical analysis. This study was initially motivated by the overarching inquiries: Are members of the \gls{icbm} community experiencing higher cancer diagnosis rates and/or encountering diagnoses at younger ages compared to national averages? Addressing these inquiries necessitates a granular examination of the registry, categorizing data by cancer type, geographical location, and specific demographic groups within the community. Consequently, our research focuses on missileers from \gls{mafb} diagnosed with \gls{nhl}. 

The choice to focus on this subgroup was made for several reasons. To begin, when national media outlets started reporting on the issue, the initial focus was on \gls{nhl}. This coverage of \gls{nhl} was due in part to the seemingly high number of cases being reported early on in the data collection process. Because there was already anecdotal evidence of a potential \gls{nhl} cluster, the reported cases had already gone through Torchlight's internal verification process. That is, although the data Torchlight collected and we subsequently utilize in this study has not been medically verified, every diagnosis input into the Torchlight database has been followed-up by a member of the Torchlight team. The follow-up includes a direct phone conversation with the military member (or the member's spouse or child in the case of deceased members) confirming all medical and service related data \citep{torchlight}. Therefore, because of the media focus and further curation of the \gls{nhl} related data within the database, there was already a lot of momentum behind concentrating our study on \gls{nhl}; and \gls{mafb} was chosen as the focal location due to the number of cases reported.

This approach of narrowing our focus aligns with the \gls{cdc} guidelines for determining cancer clusters, as they define a cancer cluster as ``a greater than expected number of the same or etiologically related cancer cases that occurs within a group of people in a geographic area over a defined period of time.'' \citep{cdc} Thus, this study aims to ascertain any statistically significant deviations with respect to the incidence rate and age at diagnosis of \gls{nhl} among \gls{mafb} missileers in comparison to national benchmarks.

Cancer concerns in relation to military service is not a new phenomenon. As more risk factors associated with military weaponry, dated building materials, and chemical exposures have been identified, the number of recent studies that look at the relationship between cancer rates and military service have grown. \cite{Zhu2009} provides one of the first well-rounded analyses of cancer incidence in the \gls{us} military population compared to the general \gls{us} population, finding that cancer incidence rates differ between the two populations with the military population having lower rates of colorectal, lung, and cervical cancers but higher rates of breast and prostate cancers. A follow-up study conducted 15 years later by \cite{Bytnar2024} found higher rates of prostate and breast cancers, particularly in 40-59 year-olds in the military population. 

In terms of research that has looked at specific career fields, \cite{Webber2021} examined the incidence rates of cancer in \gls{us} Air Force fighter aviators, finding fighters who served between 1970 and 2004 had higher odds of developing and dying from certain cancers. There are also those papers that have focused on particular types of cancer in military members, such as the work of \cite{Bytnar2021} who focused on the incidence rates of digestive cancers in military members. Additionally, in terms of relating military occupational exposure to higher cancer rates, two studies by primary author Michael Peleg consider the effects of radio frequency radiation (RFR) in relation to increased cancer incidences for military service members, with the focus of determining if exposure to RFR is a human carcinogen \citep{peleg2018,Peleg2023}. For a comprehensive overview of some of these findings and other related literature, we suggest the meta-analysis of \cite{lovejoy2024}.

Our goal in this initial analysis differs slightly from those works previously cited, as we seek to determine statistical significance of a pre-defined cancer within a specific career field at a single location over a defined period of time. Additionally, our source of cancer incidences is not a military medical affiliated database such as the Automated Central Tumor Registry (ACTUR), Veterans Affairs Central Cancer Registry (VACCR), or Defense Medical Surveillance System (DMSS). We acknowledge up front that using an outside registry has potential for missing data; however, we may also capture cases not reported through any of the above systems, such as non-retired veterans. One of the obstacles of conducting this study without the use of military databases is the lack of personnel data necessary to define the overall cohort. However, this creates an opportunity to develop new methodology in such instances where the only data in hand is for those in the population who were diagnosed. Therefore, our approach offers a method for estimating the underlying population with a corresponding assessment of how this estimation affects the overall study results. 

Examined in detail in Section \ref{sec:SIR}, the \gls{sir} is defined as the ratio of the observed over expected number of diagnoses in a given population over a specific period of time. Having partial or incomplete data on an entire cohort leads to subsequent difficulties in calculating the expected number of diagnoses. In terms of cancer related \gls{sir} analyses with this type of unknown or missing data, few studies exist that address such issues. \cite{Silocks1994} showed that when the expected number of events is subject to random variation, the beta distribution can be used for statistical inference. More recently, \cite{Becher2017} developed a method to estimate the observation times for members of a cohort with unknown follow-up data. While this aspect of the research of \cite{Becher2017} is applicable to our current study, a key difference remains in that they assume a fully defined cohort with all covariate information known, while this information must also be estimated in our analysis. Therefore, we will introduce some estimation methods based on partial data and expert knowledge to fill this gap. However, before discussing this aspect of our study, we begin with an overview of the operational environment and duties specific to the missileer career field. We then introduce the details of the data for the \gls{nhl} cases. Next, we outline the varying methodologies implemented for evaluating patterns of cancer. Finally, we apply the described methodologies to our data set to determine the existence of a cancer cluster in reference to the 2022 CDC definition, as well as in a broader context \citep{cdc}.

\section{Operational environment}\label{ops_env}
\begin{figure}[H]
\centering
\includegraphics[width=0.75\textwidth]{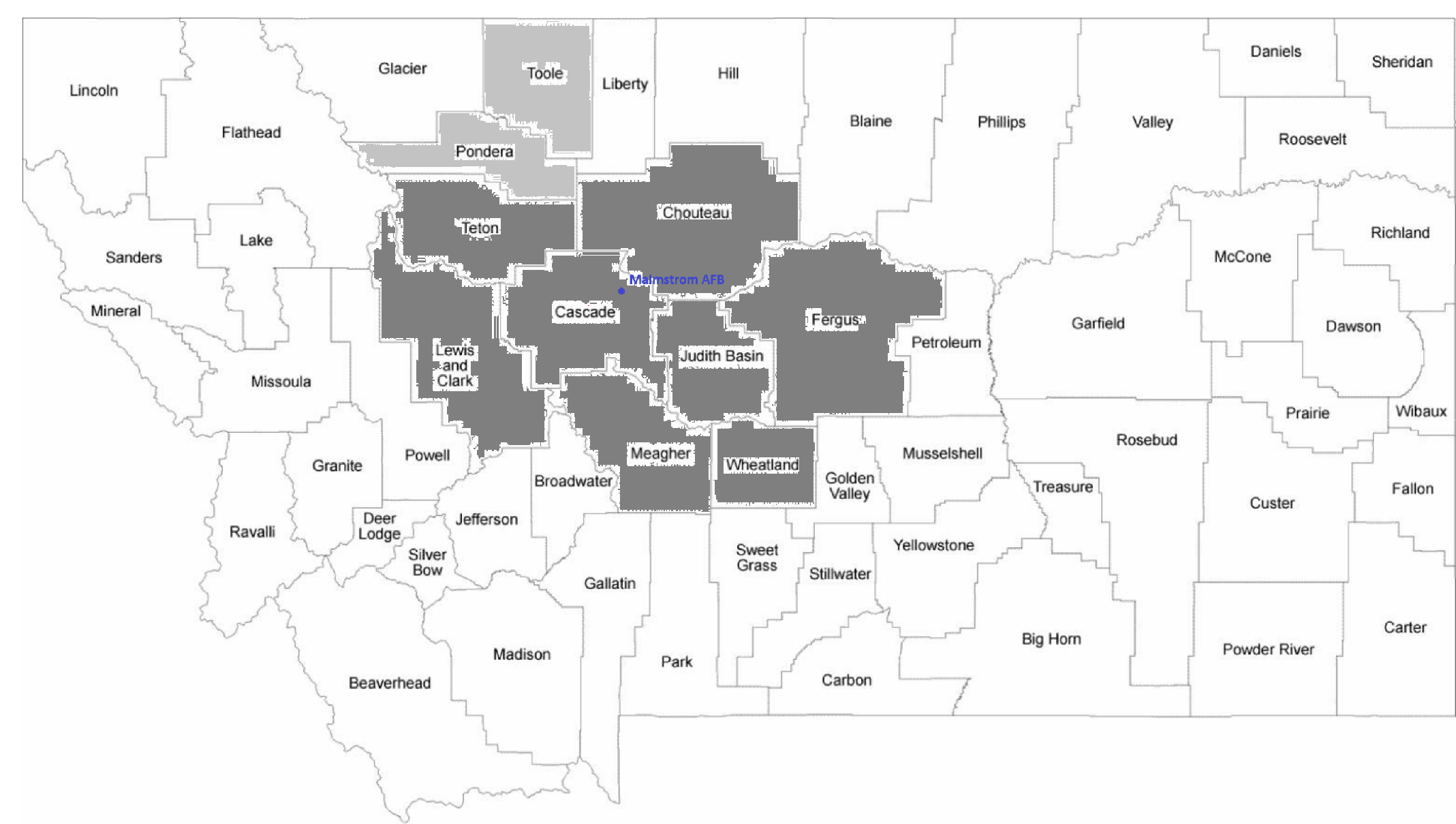}
\caption{Map of Montana counties with counties containing MAFs shaded in grey. Lighter grey indicates counties only home to 564\textsuperscript{th} MS sites.}
\label{MT_map}
\end{figure}

The missileers' unique mission within the \gls{dod} necessitates a very specific set of working conditions and everyday environment, one that has not changed since the inception of the mission in the early 1960s. To understand the potential hazards missileers face on a daily basis while performing their duties, we must examine what those duties entail. The following overview will give the average reader a better idea of what the life of a missileer looks like, to include an understanding of some of the typical terminology specific to the missile career field. Though there are three active missile bases, our description will focus on \gls{mafb}. Located in Great Falls, MT within Cascade county, \gls{mafb} is home to the 341\textsuperscript{st} \gls{mw}. Though there are now only three active squadrons, the 10\textsuperscript{th}, 12\textsuperscript{th}, and 490\textsuperscript{th} \glspl{ms}, during our years of focus there was also a fourth squadron active, the 564\textsuperscript{th} \gls{ms}. While homed at \gls{mafb}, the actual facilities (known as a \gls{maf}) where missileers perform the majority of their duties, are geographically separated and scattered throughout 10 of Montana's sprawling counties. The expanse of the 341\textsuperscript{st} \gls{mw} can be seen in Figure \ref{MT_map}.

Besides the four active squadrons, the 341\textsuperscript{st} \gls{mw} is home to two additional squadrons primarily staffed by missileers. These are the \gls{oss} and the \gls{ogv} Squadron. The primary duties of a missileer are dependent on the squadron to which they belong. If in a numbered squadron, the missileer's primary duty is to `pull alert'. This consists of spending anywhere from 24 to 72 hours on site at a \gls{maf}. The amount of time on duty at the \gls{maf} has fluctuated over the years. If on a 24 or 48 hour shift, the missileer will spend the entirety of that shift underground in the \gls{lcc}. If the schedule is a 72 hour alert, this would typically involve more than one crew manning the \gls{lcc} in shifts (where a crew is two people who man the \gls{lcc} together). Depending on the length of an alert, a missileer from a numbered squadron will have anywhere from four to eight alerts in a month. This equates to spending on average around 160 to 170 hours underground in the \gls{lcc} per month. On the other hand, when in \gls{oss} or \gls{ogv}, a missileer's alert load is a bit lightened as the primary duties of these squadrons are teaching and evaluation, respectively. The alert load in these squadrons is usually around one-third that of a missileer in a numbered squadron.

As already mentioned, a missileer's primary duty is conducted approximately 60 feet under ground behind two blast doors in a small enclosed space known as the \gls{lcc}. This capsule-shaped unit contains all the necessary equipment for the working missileers to monitor and operate the 10 \glspl{icbm} in their control. During an alert period, both blast doors are typically closed and the two crew members (missileers on duty) take shifts monitoring the in-house equipment and remotely located \glspl{icbm}. Built in the 1960s, the \glspl{lcc} have changed very little in 60 years. Due to their age and underground location, the potential exposure to health hazards associated with buildings from the 1960s are relevant risk factors today in the still operational \glspl{lcc}. These hazards include, but are not limited to: Asbestos, lead-based paint, Radon, mold, and \gls{pcbs} \citep{epa2023,cdc_lead,who2023,mayo2023,atsdr2023}.

Besides the hazards associated with a capsule's age, prolonged exposure to the array of constantly operational equipment may also be a potential threat to human health. Another consideration to human health is the air quality of a capsule. The entire \gls{lcc} is pressurized to prevent the infiltration of outside air (among other wartime threats) and the existing air is continuously circulated and filtered with the goal of maintaining air quality and preventing the buildup of harmful substances. However, the impact of internal events such as sewage build-ups, debris disturbing maintenance, and even the threat of fire could impact the quality of air within the contained system. We next discuss the specifics of the data we use in our analysis.

\section{Data}\label{data}

Before we explore the specifics of the data for the subjects diagnosed with \gls{nhl}, we provide an overview of how the data was collected and what information was provided during the data collection process. The data we are using are self-reported cancer diagnoses collected via the Torchlight Initiative, an organization that was founded to collect reports of cancer diagnoses from within the \gls{icbm} community and support those community members diagnosed with cancer. The registration form requests the following information from those voluntarily completing the form: Name, email, phone number, sex, living or deceased, date of death and age at death, bases and squadrons of service, years of service, number of alerts pulled (missileers only), age at diagnosis(es), year of diagnosis(es), details of diagnosis(es), stage/type of diagnosis, career field, permission to disclose name and illness to Congress or Investigating Agency, consent to privacy policy, and any additional information. Every registration was followed up by a member of the Torchlight team; that is, the Torchlight team authenticated the entry by speaking to the individual whose information was entered, or a spouse or child of the individual if the person was deceased. The data was de-identified before being used for statistical analysis. 

While many types of cancer have been reported through the Torchlight Initiative, there were several reasons we chose to focus on \gls{nhl} diagnoses; these included the initial number of \gls{nhl} cases the Torchlight team saw being reported, as well as the reporting presence of \gls{nhl} within mainstream media \citep{dailymail,washpost,apnews}. While \gls{nhl} is used to define many different types of lymphoma (a cancer that starts in white blood cells), all of these varying sub-types share some of the same characteristics. We next examine the provided data for the 18 cases of \gls{nhl} found in \gls{mafb} missileers. In the following sections, we examine the data for the diagnosed subjects in terms of their demographics, diagnoses, and collective time at \gls{mafb}. We consider the group as a whole, though we also examine the specifics of a smaller sub-group that served together over a shorter period of time.

\subsection{NHL cases} \label{nhl_cases}

\begin{figure}[H]
 \centering
\includegraphics[width=0.5\textwidth]{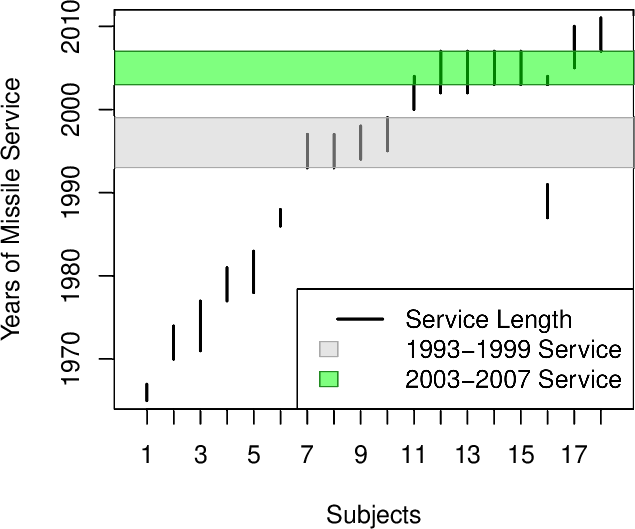}
\caption{NHL subjects years of service at \gls{mafb}}
\label{YOS}
\end{figure}

We begin by examining when each of the subjects served at \gls{mafb}. In Figure \ref{YOS}, we see the years of service at \gls{mafb} for each subject; additionally, we define two specific periods where there is considerable overlap in service dates among many of the subjects. Four of the eighteen subjects served at \gls{mafb} between 1993 and 1999, while eight of the subjects served between 2003 and 2007. 

Of note in Figure \ref{YOS} is subject 16's years of service at \gls{mafb}. Those examining Figure \ref{YOS} closely will see an earlier, longer time in service about 10 years prior to the time served within the 2003-2007 period. The later service time is most likely a return to the 341\textsuperscript{st} \gls{mw} as a more senior officer in a leadership role. In such a position, subject 16 would still have been required to perform field duties, though the frequency of those duties would be far less than a typical missileer in their first tour of duty. While no other subjects within this study displayed a similar service pattern, this career path is followed by those who pursue leadership within the career field, roughly 10-15\% of the missile force. In the following section, we briefly outline some demographics of our subjects. 

\subsection{Demographics}\label{demographics}

As outlined at the beginning of Section \ref{data}, we do not have the most detailed of demographics on our subjects; however, we do know that of the 18 in our study, only one is female. In context of \gls{af} demographics, according to a 2022 report by the \gls{us} Department of Defense, 23.4\% of \gls{af} active duty officers are female \citep{dod_dem}. We can also examine this percentage through the lens of \gls{nhl} incidence rates, which we define as the number diagnoses divided by the population time at risk. Looking at the \gls{nhl} \gls{seer} 5-year age-adjusted incidence rates from 2017 to 2021 by gender, we see that the female incidence rate is 15.5 per 100,000 as compared to the male incidence rate of 22.5 \citep{seer_1}. This means the incidence rate for males is 45.16\% higher than that for females. Thus, the combination of \gls{af} demographics and \gls{seer} incidence rates sheds more light on the gender difference in our 18 subjects. Because of the small size of our study group, we determined not to adjust rates based on gender but rather consider the rates with both males and females included.

\begin{figure}[H]
\centering
\includegraphics[width=.50\textwidth]{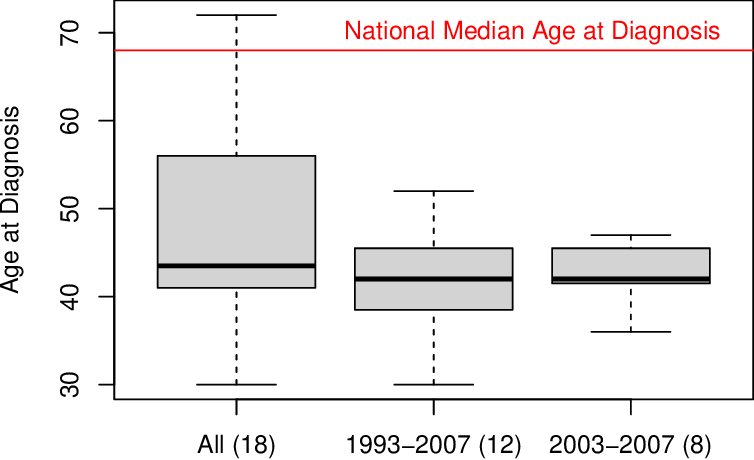}
\caption{Box plots for age at diagnosis based on different sets of MAFB NHL cancer cases from the registry}
\label{boxplots}
\end{figure}

The main additional piece of information we have is the age of the subjects upon diagnosis (and death when applicable). From this data, we can also establish how old each subject was when they were serving as missileers at \gls{mafb}. Of relevance to some of our analyses, we examine the age of diagnosis using box plots. In Figure \ref{boxplots}, we see three different box plots; the first includes all 18 subjects, the second the 12 subjects who served between 1993-2007 and the third the 8 subjects who served between 2003-2007. We see as we reduce the number included, which means including only those who served more recently, the range of our box plot decreases. However, the median age of diagnoses does not shift greatly amongst all three plots. Included for visual reference is a horizontal line annotating the national median age of diagnosis \citep{seer_1}.

\begin{figure}[H]
    \centering
    \begin{subfigure}[b]{0.42\textwidth}
        \raisebox{3.75mm}{\includegraphics[width=\textwidth]{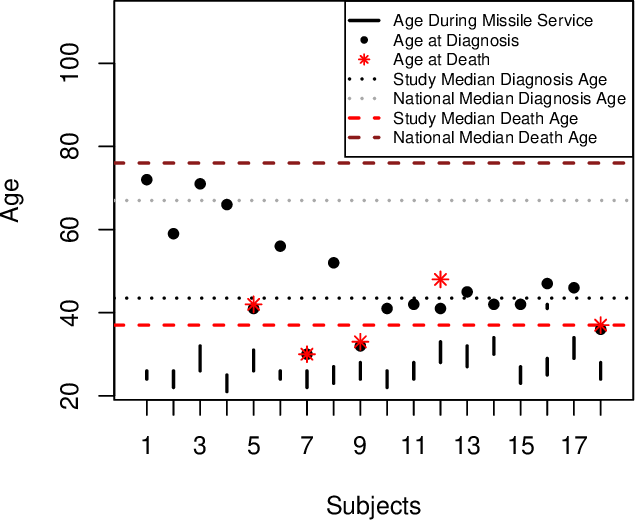}}
        \caption{Age of service, diagnosis, and death (when applicable) for all MAFB NHL missileer cases}
    \end{subfigure}
    \hspace{0.03\textwidth}
    \begin{subfigure}[b]{0.42\textwidth}
        \raisebox{-3.75mm}{\includegraphics[width=\textwidth]{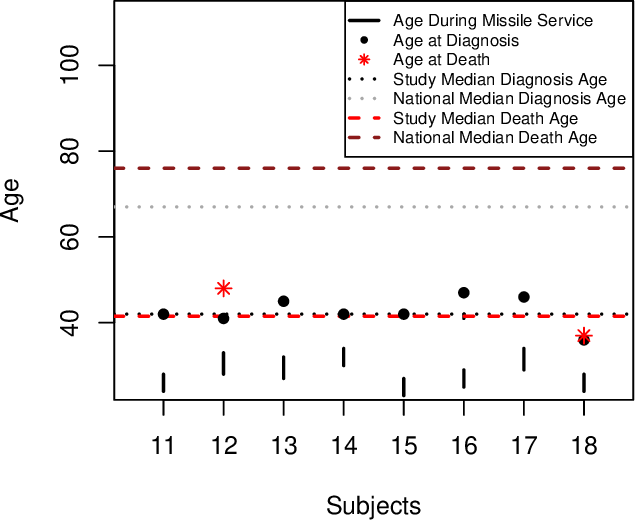}}
        \caption{Age of service, diagnosis, and death (when applicable) for 2003-2007 MAFB NHL missileer cases}
    \end{subfigure}
    \caption{Age specifics for service, diagnosis, and death}
    \label{fig:age_diag_death}
\end{figure}

We dive even further into the age at time of service, diagnosis, and death of \gls{mafb} missileer \gls{nhl} cases in Figure \ref{fig:age_diag_death}. In both Figures \ref{fig:age_diag_death}a and \ref{fig:age_diag_death}b, we see the different horizontal lines that represent the median of both diagnosis and death ages for our study population as well as national benchmarks. The main pattern seen in Figure \ref{fig:age_diag_death}a is the general decline in age of diagnosis after the first few cases, as well as the relatively young age at death for those applicable cases. We will discuss in more detail the implications of the median age at diagnosis of \gls{nhl} in the analysis section. Next we survey the service specifics of the study subjects.

\vspace{-.4in}

\subsection{Service specifics}\label{service}
\begin{figure}[H]
\centering
\includegraphics[width=.55\textwidth]{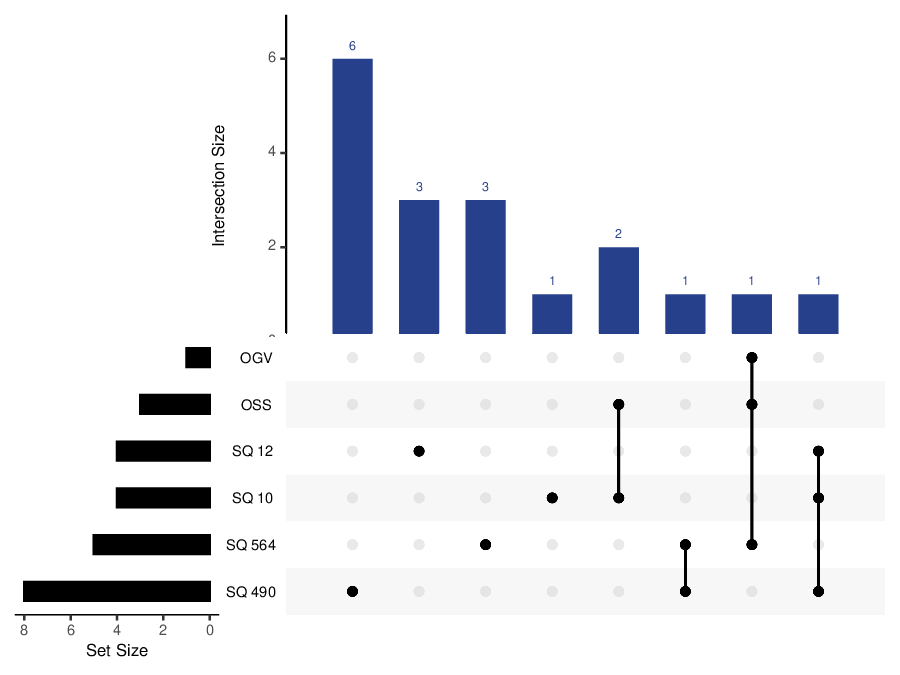}
\caption{Visualization of squadron intersections: bottom portion shows sets, histogram portion shows number in each set}
\label{upset_sq}
\end{figure}

As described in Section \ref{ops_env}, when serving in the 341\textsuperscript{st} \gls{mw} between 1965-2007, there were four active numbered squadrons and the two support squadrons to which a missileer could belong. In Figure \ref{upset_sq} we see a visualization of where each of our subjects served during their time at the 341\textsuperscript{st} \gls{mw}. The chart can be read as follows: 6 subjects served in \textit{only} the 490\textsuperscript{th} \gls{ms} (vertical barplot) while 8 subjects overall served in the 490\textsuperscript{th} \gls{ms} (horizontal barplot), with the additional two who served in the 490\textsuperscript{th} \gls{ms} also having served in either the 564\textsuperscript{th} \gls{ms} or the  10\textsuperscript{th} and 12\textsuperscript{th} \gls{ms} and so forth. The reason for examining such data follows the logic that it is fairly typical for missileers to pull alert in the numbered squadron to which they belong. Given that knowledge, exploring this data could tentatively lead us to a deeper understanding of potential sources of toxins. The main takeaway we see is that close to half of our subjects served in the 490\textsuperscript{th} \gls{ms}; and, not visible within Figure \ref{upset_sq} but upon further inspection of the data, 4 of the 8 who served in the 490\textsuperscript{th} \gls{ms} served during 2003-2007.

\begin{figure}[H]
\centering
\includegraphics[width=.65\textwidth]{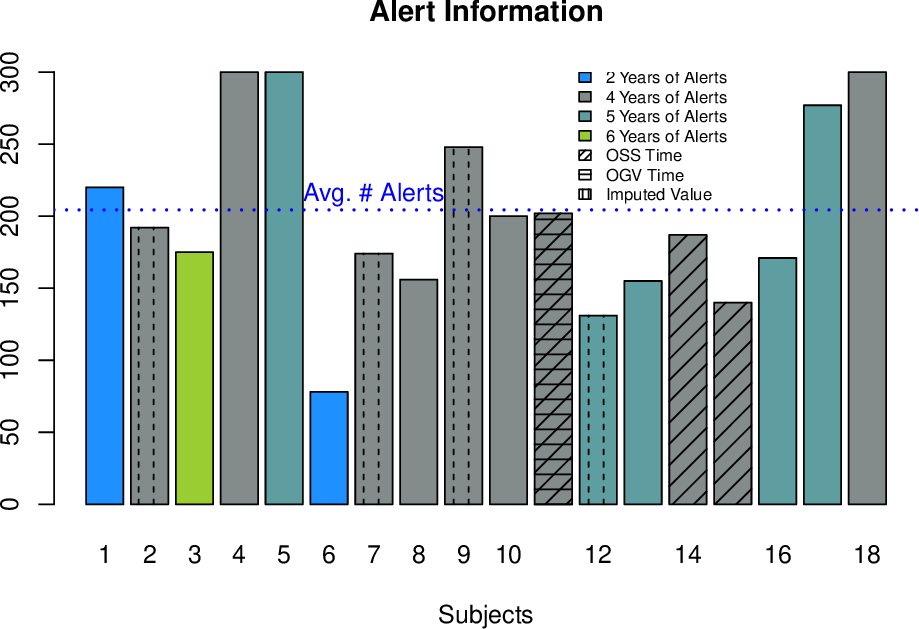}
\caption{Number of alerts by subject}
\label{num_alerts}
\end{figure}

Next, we consider the number of alerts each of our subjects pulled over their years of service at \gls{mafb}. This speaks to the amount of time each was on duty in the \gls{lcc}. In Figure \ref{num_alerts} we see the number of alerts each subject pulled to include the number of years over which these alerts took place and an indicator as to whether or not they spent time in a non-numbered squadron. Four of our eighteen subjects did not report the number of alerts they pulled (three of the four who did not report are deceased); therefore, we combined the data from the fourteen subjects who did report their alert count with another relatively small dataset we had reporting the total number of alerts twenty other missileers pulled during the 1990s. Combining these two datasets, we fit a Normal distribution to the data and imputed the values of the four missing alert counts from this Normal distribution. The subjects for whom the alert counts were imputed are indicated by dashed vertical lines in Figure \ref{num_alerts}. The blue dotted horizontal line indicating the average number of alerts (204) was taken as an average of the 14 within our dataset who reported their alert count; the fitted normal comprising those 14 and the additional 20 had a mean only slightly larger (210).  While the values are self-reported, it is typical practice for a missileer to record their alerts and the activities of those alerts for the purpose of accuracy on annual performance reports.

In general, what we glean from Figure \ref{num_alerts} is an overall concept of the amount of time each of our subjects spent in a \gls{lcc}, which could speak to the potential duration of exposure to toxins. While we have examined several aspects of the data available to us from the Torchlight registry, we note here that in this analysis the majority of these variables are not considered. Though we only outline them here to demonstrate the elements unique to the missileer career field which may speak to potential exposures, further consideration of such covariates like squadron affiliation and number of alerts are discussed in our concluding remarks in reference to future studies. Now that we have looked deeper at the service specifics of our subjects, we move on to discuss the methodologies used when conducting our analysis.

\section{Methodology}\label{sec:methodology}
In this section, we begin by describing the process employed to generate the \gls{seer} rates. We then discuss the general methodology used in evaluating incidence rates. From there we describe the method we implemented in regard to estimating the size of the underlying population. Finally, we outline a test of statistical significance that is used to determine if the age of diagnosis for our study differs significantly from the general population. However, we first note that this study was reviewed by the University of North Carolina Chapel Hill Office of Human Research Ethics which determined that this submission does not constitute human subjects research as defined by federal regulations (Reference ID 439084).

\subsection{Generating SEER rates}
In order to generate the SEER Incidence rates for NHL, we used the SEER*Stat Software \citep{seerstat_software}. Within the software, we have the following choices to generate an incidence rate table:
\begin{itemize}
    \item Database
    \item Statistic
    \item Selection
    \item Table
\end{itemize}

\subsubsection{Database}
To account for the fact that our study years span from 1962-2022, we use the SEER Research Data, 8 Registries, Nov 2023 Sub (1975-2021) database \citep{seer_database}. The numerical value of the registries (in our case 8) alludes to the number of registries from which the database is comprised. While the SEER 8 Registries database comprises the smallest number of registries of all databases (22 registries is the largest database), it covers the widest temporal range; that is, none of the larger databases account for diagnoses as far back as 1975 (the other start years are either in the early 90`s or around 2000). Though we also complete the same analyses for more restricted study periods for which the larger databases could have been applied, we chose to only use the rates from the SEER 8 Registries for consistency of analysis across study periods.

\subsubsection{Statistic}
The statistic tab allows for the choice between rates or trends with associated crude or age-adjusted options. Age-adjusted rates account for differences in age distributions across populations by weighting rates based on the age distribution of a standard population. The statistic tab also also allows for a choice of the standard population to use for comparison as well as how to define the age variable. We use age-adjusted rates with the 2000 US Std Population (19 age groups, Census P25-1130) and the age variable as Age recode with $<1$ year olds. This is standard to what SEER uses in their published rates \citep{seer_std_populations}.

\subsubsection{Selection}
Under the selection tab, the third option is the choice of Site and Morphology. In this context, site refers to the specific location in the body from which the cancer originates, and morphology refers to the cellular characteristics and behavior of the cancer. For this option, we specify the Site and Morphology to be the Site recode ICD-0-3/WHO 2008, and choose the option ``Non-Hodgkin Lymphoma''. According to SEER, this is the ``recommended variable for reporting based on the 1973-2010 SEER Research Data (November 2012 submission) and later releases.''
\citep{seer_site_recode}

\subsubsection{Table}
For the Table option, we choose the ``Age recode with $<1$ year olds'' from the Age at Diagnosis folder to be the rows of the table.

The above choices result in a table with 19 rows (one for each of the 19 age groups) and 3 columns: Rate, Count, Pop. The rate column provides the rate of NHL diagnosis for a given age group per 100,000 per year. The rate column is calculated as the proportion of the count/pop column entries. The pop column is the total population within the designated age group over the years specified in the database (1975-2021) and the count column is the number of NHL diagnoses received by the people within that age group during the years specified in the database (1975-2021). Therefore, the rate column provides us with the rates for the reference population that we use within our analysis. We next review the relevant methodology for determining statistical significance of cancer incidence rates.

\subsection{Hypothesis test based on Poisson distribution}\label{sec:hyp_test}

The following overview is taken from \cite{rothman}. Let us suppose the number of observed diagnoses $X$ follows a Poisson distribution with rate $\lambda_E$ equal to the expected number of diagnoses $(X \sim \text{Poisson}(\lambda_E))$:

\begin{equation}
    f(k|\lambda_E)=\probP(X=k)=\frac{\lambda_E^k \exp{(-\lambda_E)}}{k!}, \quad k=0,1,2,\hdots
\end{equation}
we seek to test the hypothesis:
\begin{align}\label{eq:hyp_lambda_E}
\begin{split}
       H_0 &: \lambda =\lambda_E\\
    H_1 &: \lambda > \lambda_E 
\end{split}
\end{align}
This means we are testing the null hypothesis that the true incidence rate $\lambda$ in our study population is equal to the expected rate $(\lambda_E)$ versus the alternative hypothesis that the true incidence rate is greater than the expected rate.

\subsubsection{Calculating expected rate}

In order to calculate $\lambda_E$, we must account for the amount of time each person from our study population spent in any given age group, as the SEER incidence rates are specific to five-year age groups. For the purpose of this calculation, we account for all members of the population from their entry into missile service through the end of each study period, 2022. Consider $g=1,\hdots,G$ as the number of five-year SEER age groups with which we are concerned. Then we have $r_g$ as the associated SEER NHL incidence rate for each age group $g$, and $t_g$ as the total amount of time spent in age group $g$ by the entirety of the study population.

Now let $t_{g,i}$ be the amount of time individual $i$ spends in age group $g$. For any given year, we have a total of $N_y$ individuals arriving. Therefore, $t_{g,y}$ accounts for the amount of time spent in study group $g$ by those arriving in year $y$:

\begin{equation}\label{eq:time_person}
    t_{g,y} = \sum_{i=1}^{N_y}{t_{g,i}}
\end{equation}

From there, we total the amount of time spent in group $g$ by adding up $t_{g,y}$ for every year of our study period, $y=1\hdots,Y$:

\begin{equation}\label{eq:time_yr}
    t_g = \sum_{y=1}^Y{t_{g,y}}
\end{equation}

Then the final calculation for $\lambda_E$ is given by:

\begin{equation}\label{eq:lambda_E}
    \lambda_E = \sum_{g=1}^G{r_gt_g}
\end{equation}

\subsubsection{Evaluating hypothesis test}

After calculating $\lambda_E$ using Equations \ref{eq:lambda_E}, we then consider several methods to evaluate the hypothesis test given in Equation \ref{eq:hyp_lambda_E}. Using a large-sample statistic, our Z-score statistic is 

\begin{equation}
    Z_{score} = \frac{X-\lambda_E}{\sqrt{\lambda_E}}
\end{equation} 

where our p-value for Equation \ref{eq:hyp_lambda_E} is given by:

\begin{equation}
    p_z = 1-\Phi(Z_{score})
\end{equation}

with $\Phi$ indicating the \gls{cdf} of a standard normal distribution.

This is based on the asymptotic normality of the MLE $\hat{\lambda}_E=X$, where $X$ is the number of observed cancer cases. Another approach to evaluating Equation \ref{eq:hyp_lambda_E} in the absence of a large sample size is to directly use the Poisson distribution. A p-value based on the Poisson distribution for Equation \ref{eq:hyp_lambda_E} is given by:

\begin{equation}
    p_e=\probP(X \geq k|\lambda_E) = 1 - \probP(X \leq k-1|\lambda_E)=1-\sum_{x=0}^{k-1}{\frac{\lambda_E^x \exp{(-\lambda_E)}}{x!}}
\end{equation}

$p_e$ is known as the ``exact" p-value and is often more appropriate with small samples as the discreteness of the Poisson distribution becomes more significant in such cases, and $p_z$ may not accurately reflect the true distribution of the observed counts (as it relies on asymptotic results).

The last method we present on evaluating Equation \ref{eq:hyp_lambda_E} is the mid-p-value approach. Again relying on the Poisson distribution, the mid-p-value calculation attempts to refine the p-value calculation of $p_e$ by providing a more accurate measure of statistical significance. It does this by averaging the probability of observing exactly $k$ events and the cumulative probability of observing more extreme events. That is, for our one-tailed hypothesis test, the mid-p-value is given by:

\begin{equation}\label{eq:upper_mid_p}
    p_{m_u} = \frac{1}{2}\probP(X=k|\lambda_E) + \probP(X > k | \lambda_E)
\end{equation}

The mid-p-value is typically less conservative but more powerful than the exact p-value. That is, using the mid-p-value means a potential slight increase in the Type I error rate in exchange for more frequent rejections of $H_0$ when it is indeed false \citep{Hirji2006}.

\subsubsection{Examining the Standardized Incidence Ratio (SIR)}\label{sec:SIR}

Associated with the hypothesis test given in Equation \ref{eq:hyp_lambda_E} is the standardized incidence ratio (SIR). The SIR is given as the ratio of the true rate parameter $\lambda$ and the expected rate based on the reference population, $\lambda_E$. When we evaluate the MLE of this ratio, we end up with the proportion of observed over expected diagnoses in a given population over a specified period of time:

\begin{equation}
    \hat{SIR} = \frac{X}{\lambda_E}
\end{equation}

The approximate Wald confidence interval based on the normality assumption for $X$ as $\lambda_E$ increases is given by \citep{rothman}:

\begin{equation}
    100(1-\alpha)\% \text{CI} = \exp{\big(\ln{(\hat{SIR})}\pm Z_{\frac{\alpha}{2}}\times \text{SE}\big)}
\end{equation}

where $\text{SE}=\sqrt{\frac{1}{X}}$ is the standard error of $\ln{(\hat{SIR})}$. If $\hat{SIR}>1.0$ and the resulting confidence interval does not contain $1.0$, then the conclusion can be drawn that there is a nonrandom excess of NHL diagnoses among MAFB missileers.

We can also derive a confidence intervals associated with the exact p-value and mid-p-value methods. There are several approaches associated with exact confidence intervals; for this analysis, we use the method introduced by \cite{garwood1936}. Still a relevant and popular CI estimation approach included in almost all statistical software packages for such analyses, this interval is given as \citep{pradhan2021}:
\begin{equation}
    100(1-\alpha)\%\text{CI} = \left(\frac{\frac{1}{2}\chi^2_{2x,\alpha/2}}{\lambda_E},\frac{\frac{1}{2}\chi^2_{2(x+1),1-\alpha/2}}{\lambda_E}\right)
\end{equation}
where $x$ is the observed number of cases.

Finally, to calculate the CI associated with the mid-p-value technique, we re-frame the Poisson distribution with respect to the SIR. That is, under the assumption that $SIR=\frac{\lambda}{\lambda_E}=1.0$, we can rewrite the \gls{pmf} of the Poisson as:

\begin{equation}
    \probP(X=k)=\frac{(SIR\times \lambda_E)^k \exp{(-SIR \times \lambda_E)}}{k!}
\end{equation}

We have already stated the upper mid-p-value in Equation \ref{eq:upper_mid_p}; the lower mid-p-value is given as $p_{m_l}=\frac{1}{2}\probP(X=k|\lambda_E)+\probP(X <k |\lambda_E)$. In order to get a two-sided $(1-\alpha)$-level mid-p confidence interval for $\hat{SIR}$, we take the lower limit to be the value for $SIR$ for which $p_{m_u}=\alpha/2$, and take the upper limit to be the value for $SIR$ for which $p_{m_l}=\alpha/2$ \citep{rothman}. 

\subsection{Testing diagnosis age discrepancy}\label{age_test}
While the previously defined methods may inherently speak to a discrepancy in terms of \gls{nhl} diagnosis age between \gls{mafb} missileers and the national values, we seek to test this difference directly. \gls{seer} reports the national median age at diagnosis and death for a variety of cancers, to include \gls{nhl} \citep{seer_explorer}. The use of the median as the reported statistic is likely due to the skewed nature of cancer diagnoses. Therefore, in order to test if the median age at diagnosis for the \gls{mafb} missileer population differs from the national median, we use a non-parametric test. The test we employ is known as the Sign Test and is non-parametric as it does not make any distributional assumptions about the data (e.g. normality). It merely assesses whether there is a consistent tendency for the observed median of our study population $(\eta)$ to be on one side of the hypothesized median, the national median $(\eta_0)$, based on the number of observations above and below it. As described by \cite{hollander}, we can see the null and alternative hypotheses in Equation \ref{eq:hyp}, where our alternative hypothesis is that the median of our study population is less than the national median. 

\begin{equation}\label{eq:hyp}
\begin{aligned}
    H_0 &: \eta = \eta_0 \\
    H_1 &: \eta < \eta_0
\end{aligned}
\end{equation}

The test works as follows: for each data point $Z_i, i=1,\hdots,n$ (which we take to be the subjects age at diagnosis), we define an indicator variable $\psi_i, i=1,\hdots,n$ such that 
\begin{equation}\label{eq_indicator}
  \psi_i=\begin{cases}
    1, & \text{ if } Z_i>\eta_0\\
    0, & \text{ if } Z_i<\eta_0
  \end{cases}
\end{equation}

We then define the sign statistic $B$ as:

\begin{equation}\label{eq:Bval}
    B=\sum_{i=1}^n \psi_i
\end{equation}

From here, at the chosen $\alpha$ level of significance we reject $H_0$ if $B \leq n-b_{\alpha,1/2}$, where the constant $b_{\alpha,1/2}$ is chosen to make the type I error probability equal to $\alpha$ and is the upper $\alpha$th percentile point for the binomial distribution with sample size $n$ and $p=\frac{1}{2}$. We can also find the associated p-value of our test by calculating: 

\begin{equation}\label{eq:pvalue}
    F(B|n,p) = \text{P}(X \leq B) = \sum_{i=0}^{\lfloor B \rfloor} \binom{n}{i} p^i (1-p)^{n-i}
\end{equation}

 Additionally, based on our one-sided hypothesis test, we can calculate a one-sided confidence interval regarding the median $\eta$ of our study population; that is, we estimate an assertion, with a specified level of confidence, that $\eta$ is no larger than some upper confidence bound based on the sample data. We next examine how to estimate the age and size of the underlying population.

\subsection{Estimation of unknown quantities}

We only have information on the missileers who served at \gls{mafb} who were diagnosed with NHL. Therefore, in order to proceed with any type of analysis (specifically in order to calculate the expected number of \gls{mafb} missileers diagnosed with NHL), we must make some assumptions about the size of the population and the age of the individuals within the population when they began missile service. As previously noted, this need for estimation differs from the work done by \cite{Becher2017} where this particular data was assumed known. To account for these unknowns, our approach suggests assigning a distributional form to the number of personnel joining the underlying population every year as well as assigning the start age each member of the population began missile service from a previously defined distribution. 

From historical arrival rates provided by personnelists from \gls{mafb} we generate the number of missileers arriving each year from a normal distribution with mean $\mu_y=75$ and standard deviation $\sigma_y=7$. However, because of political and military events throughout the decades, we are aware of various anomalies and changes in the rate of arrival. Therefore, we run several simulations to quantify the uncertainty associated with the choice of $\mu_y$. Within these simulations, we vary the mean of this normal distribution, which allows us to create confidence intervals surrounding the expected number of diagnoses.

In a similar manner, we estimate the parameters of the normal distribution for the start age of missile service based on the start age of our subjects diagnosed with NHL. While we have a relatively small sample size of these start ages, the resulting mean and standard deviation $(\hat{\mu}_a,\hat{\sigma}_a)=(24.67, 2.47)$ concur with our subject matter expert's general knowledge of the start age for missileers (based on the average commissioning age, missile qualification training time, and additional special duty or re-training factors). Therefore, when calculating the expected number of MAFB missileers diagnosed with NHL, the start age for each individual within the population is generated from the stated normal distribution such that:
\begin{equation}\label{eq:age_distro}
    A=\max{(21,\omega)}, \quad \omega \sim N(24.67,2.47)
\end{equation}
where we account for 21 being the youngest age for a commissioned officer based on the education requirements. The next section reviews the results from applying the methods just described.

\section{Results}\label{sec:NHL_results}

We begin by outlining the simulation conducted to estimate the expected number of diagnoses in our population. From there, we define the specific temporal periods over which we run our analyses. We then provide the results for the non-parametric test conducted.

\subsection{Simulation for the expected number of diagnoses}\label{sec:sim}

In order to evaluate the influence the arrival rate of missileers in a given year has on the expected number of NHL diagnoses for our population, we conduct a simulation. We consider setting the mean of our normal distribution for the number of missileers arriving at MAFB to the following values $\boldsymbol{\mu}_y = \{50,55,60,\hdots,90,95,100\}$, indexed by $m=1,\hdots,M$. Then, for each value of $\mu_{y,m}$ in $m=1,\hdots,M$, we calculate the expected number of diagnoses, $\lambda_E$, $R=1,000$ times. We outline the simulation in Algorithm \ref{alg:lambda_E}.

While Algorithm \ref{alg:lambda_E} may appear complicated, it essentially shows the replicated process of calculating Equations \ref{eq:time_person}, \ref{eq:time_yr}, \ref{eq:lambda_E} $R$ times for each $m$ value of $\boldsymbol{\mu}_y$. The results for the study period 1962-2022 are given in Table \ref{tab:mean_ci_table} where the mean and associated 95\% confidence intervals are calculated for each of the different simulated values for $\mu_{y,m}$ as well as overall across all values of $\mu_y$. From the results we see that the overall mean concurs with the mean of the simulation for the suggested value of $\mu_y=75$ with just a slightly wider confidence interval.

In general, what the results in Table \ref{tab:mean_ci_table} suggest is that the rate of missileer arrival at \gls{mafb} has a large effect on the expected number of NHL diagnoses among the population. This is precisely because the magnitude of person-time grows with the inclusion of more individuals within the population. In other words, as we increase the mean of the arrival rate distribution by five, we see an increase of about two expected NHL diagnoses. Moving forward, we use the overall mean of 29.64 as the expected number of NHL cases among \gls{mafb} missileers who served between 1962-2022.

\begin{algorithm}[H]
\DontPrintSemicolon 
\KwIn{M - number of different $\mu_y$ values, R - number of replications for each $\mu_y$ value, Y - number of years in study period, G - number of five-year age groups, $\boldsymbol{\mu}_y$ - vector of values for the mean of the normal distribution for the number of missileers arriving at MAFB in a given year, $A$ - distribution for start age of missileers found in Equation \ref{eq:age_distro}}
\KwOut{$\boldsymbol{\Lambda}_{E,R\times M}$ - a matrix of $\lambda_E$ values with dimensions $R \times M$}
\BlankLine
\For{$m \in 1,2,\hdots,M$}{
   set value for $\mu_{y,m}$\;
  \For{$r \in 1,2,\hdots,R$}{
        \For{$y \in 1,2, \hdots, Y$}{
            generate number of people arriving in year $y$: $N_y \sim N(\mu_{y,m},\sigma_y)$\;
            \For{$i \in 1,2,\hdots,N_y$}{
                generate start age for individual $i$, $A_i \sim A$\;
                \For{$g \in 1,2,\hdots,G$}{
                    calculate $t_{g,i}$ - the amount of time individual $i$ spent in age group $g$\;
                } 
            } 
            \For{$g \in 1,2,\hdots,G$}{
                   calculate $t_{g,y}$ - the amount of time spent in age group $g$ by those arriving in year $y$ \;
                } 
        } 
        \For{$g \in 1,2,\hdots,G$}{
                   calculate $t_{g}$ - the amount of time spent in age group $g$ by the entire study population\;
                } 
    calculate $\lambda_{E_{m,r}}$ - the expected number of NHL diagnoses among MAFB missileers for simulation run $r$ using $\mu_{y,m}$\;
  } 
} 
\Return{Result: $\boldsymbol{\Lambda}_{E,R\times M}$}\;
\caption{Simulation for the expected number of NHL diagnoses among MAFB missileers}\label{alg:lambda_E}
\end{algorithm}

\begin{table}[H]
\centering
\small
\begin{tabular}{c|c}
$\mu_y$ & $\lambda_E$ (95\% CI) \\
\hline
$50$ & 19.77 (19.74, 19.80) \\ 
$55$ & 21.75 (21.72, 21.78) \\ 
$60$ & 23.71 (23.68, 23.75) \\ 
$65$ & 25.68 (25.65, 25.71) \\ 
$70$ & 27.71 (27.68, 27.74) \\ 
$\textit{75}$ & \textit{29.64 (29.61, 29.68)} \\ 
$80$ & 31.62 (31.58, 31.66) \\
$85$ & 33.58 (33.55, 33.61) \\ 
$90$ & 35.56 (35.53, 35.59) \\ 
$95$ & 37.55 (37.51, 37.58) \\ 
$100$ & 39.61 (39.54, 39.51) \\ \hline
\textbf{Overall} &  \textbf{29.64 (29.52, 29.76)} \\ \hline
\end{tabular}
\caption{Means and 95\% Confidence Intervals for each value of $\mu_{y,m}$}
\label{tab:mean_ci_table}
\end{table}

\subsection{Results for multiple study periods}

For our first analysis, we also want to evaluate the hypotheses given by Equation \ref{eq:hyp_lambda_E} during shorter study periods. To that end, we also calculate $\lambda_E$ for the study periods 1980-2022, 1990-2022, 2000-2022; however, in order to capture the majority of those who served during those years, we start each study period four years earlier, e.g. 1977, 1987, 1997, respectively. The reason for the gradual truncation in the time period is to ascertain if there is a temporal component to the rate of expected diagnoses. We plan to eventually consider a survival analysis which would allow us to include time as a covariate; but for now, we apply the methods from Section \ref{sec:methodology} separately for each time period and adjust the results accordingly for multiple testing using a Bonferroni correction \citep{Hochberg1987}. The results from applying the methods in Section \ref{sec:methodology}, including the calculated expected number of diagnoses for each study period $(\lambda_E)$, can be seen in Table \ref{tab:sir_results}. As stated previously, we use the overall mean value from our expected number of diagnoses simulation for $\lambda_E$ for each of the different study periods. 
\begin{table}[H]
\centering
\footnotesize
\begin{tabular}{c|c|c|c|c|c}
Study Period &$X=$ Observed & $\lambda_E$, 95\% CI & p-value & SIR & SIR $(1-\alpha_B)\cdot100$\% CI \\ \hline 
\multirow{3}{*}{1962-2022} & \multirow{3}{*}{18} & \multirow{3}{*}{$29.64, (29.52,29.76)$} & $p_z = 0.9837$& \multirow{3}{*}{$0.6073$} &$(0.3581,1.0300)_z$\\
& & & $p_e=0.9914$ & & $(0.3087,1.0666)_{e}$\\
& & & $p_m=0.9837$ & & $(0.3187,1.0476)_m$\\ \hline

\multirow{3}{*}{1980-2022} & \multirow{3}{*}{15} & \multirow{3}{*}{$10.16, (10.12,10.20)$} & $p_z = 0.0645$ & \multirow{3}{*}{$1.4764$} &$(0.8277,2.6335)_z$\\
& & & $p_e=0.0921$ & & $(0.6961,2.7278)_{e}$\\
& & & $p_m=0.0733$ & & $(0.7235,2.6720)_m$\\ \hline

\multirow{3}{*}{1990-2022} & \multirow{3}{*}{12} & \multirow{3}{*}{$4.42, (4.40,4.44)$} & $p_z = 1.56e-04$ & \multirow{3}{*}{$2.7149$} &$(1.4215,5.1852)_z$\\
& & & $p_e=2.08e-03$ & & $(1.1519,5.3677)_{e}$\\
& & & $p_m=1.38e-03$ & & $(1.2097,5.2381)_m$\\ \hline

\multirow{3}{*}{2000-2022} & \multirow{3}{*}{8} & \multirow{3}{*}{$1.679, (1.672,1.685)$} & $p_z = 5.35e-07$ & \multirow{3}{*}{$4.7647$} &$(2.1571,10.5244)_z$\\
& & & $p_e=3.56e-04$ & & $(1.5919,10.8427)_{e}$\\
& & & $p_m=2.11e-04$ & & $(1.7168,10.4944)_m$\\ \hline
\end{tabular}
\caption{Results for different study periods with various p-value and SIR CI calculations}
\label{tab:sir_results}
\end{table}

From Table \ref{tab:sir_results} we evaluate the various p-values, comparing them to our Bonferroni corrected alpha-level $\alpha_B = \frac{\alpha}{S}=\frac{0.05}{4}=0.0125$ where $\alpha$ is our initially chosen level of significance and $S$ is the number of study groups over which we are testing. Additionally, we can examine the SIR confidence intervals. In comparing the p-values to $\alpha_B$, we see that we reject $H_0$ from Equation \ref{eq:hyp_lambda_E} for the later two study periods but not the first two study periods; similarly, the SIR CIs for the later two study periods do not contain $1.0$, which suggests that the rate of observed NHL diagnoses among missileers at MAFB is greater than the expected number of NHL diagnoses during those study time periods.

In general, these results suggest that those entering missile service at MAFB in later years are being diagnosed with NHL at a higher rate than their predecessors. Recall that the value for $\lambda_E$ was essentially calculated under the assumption that an average of 75 new missileers arrived at MAFB every year (though in reality, we took an overall average from our simulation). We might be interested in asking what happens in the later study periods if we assume a higher mean rate of arrival? To answer this question, we use $\lambda_E$ associated with the 95\%ile of all our simulated values and examine the results (essentially this corresponds to assuming a mean rate of 100 missileers arriving per year). The results using the 95\%ile $\lambda_E$ values can be found in Table \ref{tab:sir_results_smaller}. 

\begin{table}[H]
\centering
\footnotesize
\begin{tabular}{c|c|c|c|c|c}
Study Period &$X=$ Observed & $\lambda_E$ (95\%ile value) & p-value & SIR & SIR $(1-\alpha_B)\cdot 100$\% CI \\ \hline 

\multirow{3}{*}{1990-2022} & \multirow{3}{*}{12} & \multirow{3}{*}{$5.8808$} & $p_z = 5.81e-03$ & \multirow{3}{*}{$2.0405$} &$(1.0684,3.8971)_z$\\
& & & $p_e=1.75e-02$ & & $(0.8658,4.0344)_{e}$\\
& & & $p_m=1.26e-02$ & & $(0.9092,3.9370)_m$\\ \hline

\multirow{3}{*}{2000-2022} & \multirow{3}{*}{8} & \multirow{3}{*}{$2.2348$} & $p_z = 5.75e-05$ & \multirow{3}{*}{$3.5797$} &$(1.6207,7.9070)_z$\\
& & & $p_e=2.18e-03$ & & $(1.1960,8.1461)_{e}$\\
& & & $p_m=1.35e-03$ & & $(1.2899,7.8844)_m$\\ \hline
\end{tabular}
\caption{Results for later study periods assuming 95\%ile value for $\lambda_E$ with various p-value and SIR CI calculations}
\label{tab:sir_results_smaller}
\end{table}
\vspace{-.1in}

While only one p-value and SIR show statistical significance for the 1990-2022 study period, all results are still statistically significant for the last study period. Thus, even when considering a much higher rate of arrival for missileers at MAFB (which in turn means a larger study population), we still find a higher rate of NHL diagnoses than expected. Before moving forward, we note that assuming a rate of 100 missileers arriving per year is extremely conservative and the results in Table \ref{tab:sir_results_smaller} should be considered with that conservative assumption in mind.

\subsection{Sign Test results}
Finally, we examine the results of the Sign Test with respect to the four study periods. Recalling from Section \ref{age_test}, we are comparing the median age at diagnosis of \gls{nhl} of our study group with the national median age at diagnosis. Our alternative hypothesis is that the median age of our study group is less than the national median age, which \gls{seer} lists as 68 \citep{seer_explorer}.

\begin{table}[H]
\centering
\small
\begin{tabular}{c|c|c|c}
\hline
Study Period & Estimated Median & p-value & Upper 95\% Confidence Bound\\ \hline
1962-2022 & 43.5 & $6.56e-04$ & 51.74 \\
1980-2022 & 42.0 & $3.05e-04$ & 46.09 \\
1990-2022 & 42.0 & $2.44e-04$ & 45.20 \\
2000-2022 & 42.0 & $3.91e-03$ & 45.68 \\ \hline
\end{tabular}
\caption{Sign Test results across study periods}\label{tab:sign_results}
\end{table}
\vspace{-.1in}

 The results of all four tests can be seen in Table \ref{tab:sign_results} where the estimated means, p-values, and upper 95\% confidence bounds are displayed. Comparing all four p-values to $\alpha_B$, we see that we reject the null hypothesis for each test. Therefore, we have sufficient evidence at the chosen $\alpha$ value that the median age of \gls{nhl} diagnosis for \gls{mafb} missileers is less than the national median age at diagnosis. We next discuss the implications of these results, the limitations identified throughout the study, and suggestions for future work.

 \section{Conclusions and future work}

 In this study, our main goal was to determine if any statistically significant deviations exist with respect to the incidence rate and age at diagnosis of \gls{nhl} among \gls{mafb} missileers in comparison with national benchmarks. We found that when evaluating the expected rate via simulation, increasing the mean number of missileers arriving at \gls{mafb} in a given year by five resulted in an increase of approximately two more expected number of \gls{nhl} diagnoses among \gls{mafb} missileers. Applying the Poisson distribution and evaluating the corresponding hypothesis test and examining the \gls{sir}s across four study periods showed evidence in later study periods of deviations from expected rates. The Sign Test highlighted a disparity between the study population and the national median age of diagnosis. 
 
 These tests were not without their limitations. First, the data that was utilized was self-reported and was not verified in the medical sense, though verification of all diagnoses and deaths were internally confirmed by the \cite{torchlight}. In terms of data, we also had a fairly small sample size, which constrains the power of our analyses. Additionally, we implemented some estimation techniques new to this type of study that added a layer of uncertainty to the predefined statistical tests. While this step added uncertainty, it also offers a new approach to conducting such analysis in the face of incomplete/unknown data. 

 On this note, we mention a few details of the estimation process that affect the overall analyses and should be considered in future studies. To begin, when calculating the person-time within Algorithm \ref{alg:lambda_E}, we did not consider early censoring due to death of those included within the underlying population. This means we assumed every member of the study population lived through the end of the study time period. This was done purposefully to provide a conservative expected rate; however, it is an unrealistic assumption given the start year of the overall study period. Adjusting the person-time calculation for such considerations was a primary contribution made by \cite{Becher2017}. Additionally, we did not allow for any latency period, meaning we assumed person-time accumulated from the first year of missile service. Again this is somewhat unrealistic given the potential for occupational exposure to a carcinogen would typically require such a latency period. Therefore, future studies should consider implementing such additions as well as developing the uncertainty quantification elements included in this study into more formal methodology.

 Though we focus on \gls{nhl} in missileers from \gls{mafb}, there is a wealth of data contained within the \cite{torchlight} registry that requires examination. Future studies should implement such methods as those applied within this analysis; however, a survival analysis, potentially applying a Cox Proportional Hazard Model, would also be warranted. This would allow for the inclusion of covariates such as age, year of missile service entry, gender, race/ethnicity, squadron affiliations, number of alerts pulled, and number of years at the missile wing etc. Because a Cox Proportional Hazard Model requires the covariate information for the entirety of the population being studied \citep{Cox1984}, we suggest a precursor analysis in the form of a discrete event simulation in order to create a synthetic underlying population in which all the covariates are assigned and during which early censorship can also be applied.

 While the approach taken in this work is unconventional in that we performed a cohort-style analysis without complete cohort knowledge, it also takes a step to fill such gaps where similar situations might exist but some statistical analysis is still desired. With a conservative approach, to include a conservative correction factor, we nonetheless found statistical significance and temporal trends within the overall analysis. Ultimately, these results suggest potential underlying risk factors or exposures unique to this population, particularly those entering in later decades, and necessitates further epidemiological scrutiny.


\bibliography{bib}





\newpage
\printglossaries
\end{document}